\documentclass[preprint, review, 3p, authoryear, times]{elsarticle}

\usepackage{todonotes}
\usepackage{subcaption} 

\usepackage{amsmath}  
\usepackage{capt-of}  
\usepackage{graphicx}
\usepackage{url}
\usepackage{color, soul}
\usepackage{bm}
\usepackage{amsmath} 
\usepackage{lineno}

\begin{document}

\title{Empowering Cognitive Digital Twins with Generative Foundation Models: Developing a Low-Carbon Integrated Freight Transportation System}


\author[utk]{Xueping Li}
\author[ornl]{Haowen Xu \corref{cor1}} 
\author[utk]{Jose Tupayachi} 
\author[ornl]{Olufemi Omitaomu} 
\author[utk]{Xudong	Wang} 

\cortext[cor1]{Corresponding author.}

\affiliation[utk]{organization={University of Tennessee, Knoxville},
            addressline={2233 Volunteer Bl
            vd.}, 
            city={Knoxville},
            postcode={37996}, 
            state={TN},
            country={USA}}

\affiliation[ornl]{organization={Oak Ridge National Laboratory},
            addressline={1 Bethel Valley Rd}, 
            city={Oak Ridge},
            postcode={37830}, 
            state={TN},
            country={USA}} 

\begin{frontmatter}
\begin{abstract}
Effective monitoring and operation of freight transportation system is crucial for advancing sustainable, low-carbon economies. Traditional approaches, which rely on discrete simulations of each mode using single-modal data and simulation, are inadequate for optimizing intermodal and synchromodal systems in a holistic manner. These systems involve complex, interconnected processes that impact shipping time, costs, emissions, and socio-economic factors. Developing digital twins to enable real-time situational awareness, predictive analytics, and optimization of urban logistics systems often demands extensive efforts in knowledge discovery, as well as significant time for integrating various datasets, coupling multi-domain simulations, and developing key software components. Recent advancements in generative AI present new opportunities to create foundational models for scientific research, significantly streamlining the development of digital twins. These models extend traditional digital twins' capabilities by automating the processes of knowledge discovery, data integration, and management to generate innovative simulation and optimization solutions. They also promote autonomous workflows for data engineering, analytics, and software development. This vision paper proposes an innovative paradigm that leverages the power of generative foundation models to enhance digital twins for urban research and operations. Using the decarbonization of the integrated freight transportation as a case study, we propose a conceptual framework that employs recent generative AI advancements, specifically transformer-based language models, to enhance an existing urban digital twin through the development of foundation models. We present preliminary results and share our vision for a more intelligent, autonomous, general-purpose digital twins for optimizing integrate freight transportation system ranged from from multimodal to synchromodal paradigms.

\end{abstract}


\end{frontmatter}



\section{INTRODUCTION}
\label{Introduction}
The global freight transportation sector, encompassing road, rail, and waterways, contributes significantly to greenhouse gas (GHG) emissions, accounting for about 11\% of global emissions \citep{international2012co2}. In the United States alone, transportation was responsible for 38\% of energy-related emissions in 2021, totaling 4.6 billion metric tons of CO$_{2}$ \citep{shirley2022emissions}. These figures highlight the sector’s substantial impact on climate change and the urgent need for sustainable approaches to mitigate emissions. Recent disruptions, notably the COVID-19 pandemic, exposed vulnerabilities in freight networks, underscoring the need for resilience and adaptability in logistics \citep{golan2020trends, dubey2022frugal}. Ensuring robust and efficient supply chains has become essential for social, economic, and environmental sustainability. This requires rethinking freight transportation design and management to better cope with future disruptions, during the process the freight transportation paradigm evolves overtime from the unimodal to multimodal, then to intermodal \citep{rossolov2017research}. In the past decade, the intermodal freight transportation paradigm has become foundational in modern supply chains. It involves moving goods between origin and destination through multiple transport modes, like trucks, trains, and ships, without handling the cargo during transitions \citep{turbaningsih2022multimodal, kengpol2014development}. By leveraging the strengths of each mode, intermodal freight optimizes efficiency, reduces costs, and minimizes environmental impact, offering significant potential to reduce carbon footprints through improved route planning, shipment consolidation, and the use of more energy-efficient modes, such as rail and waterways \citep{demir2019green}. In very recent years, the concept of synchromodal freight system has been proposed on top of the intermodal freight system with the extended ability for making real-time decisions \citep{giusti2019synchromodal}.

\begin{figure*}[hbt!]    \includegraphics[width=1\textwidth]{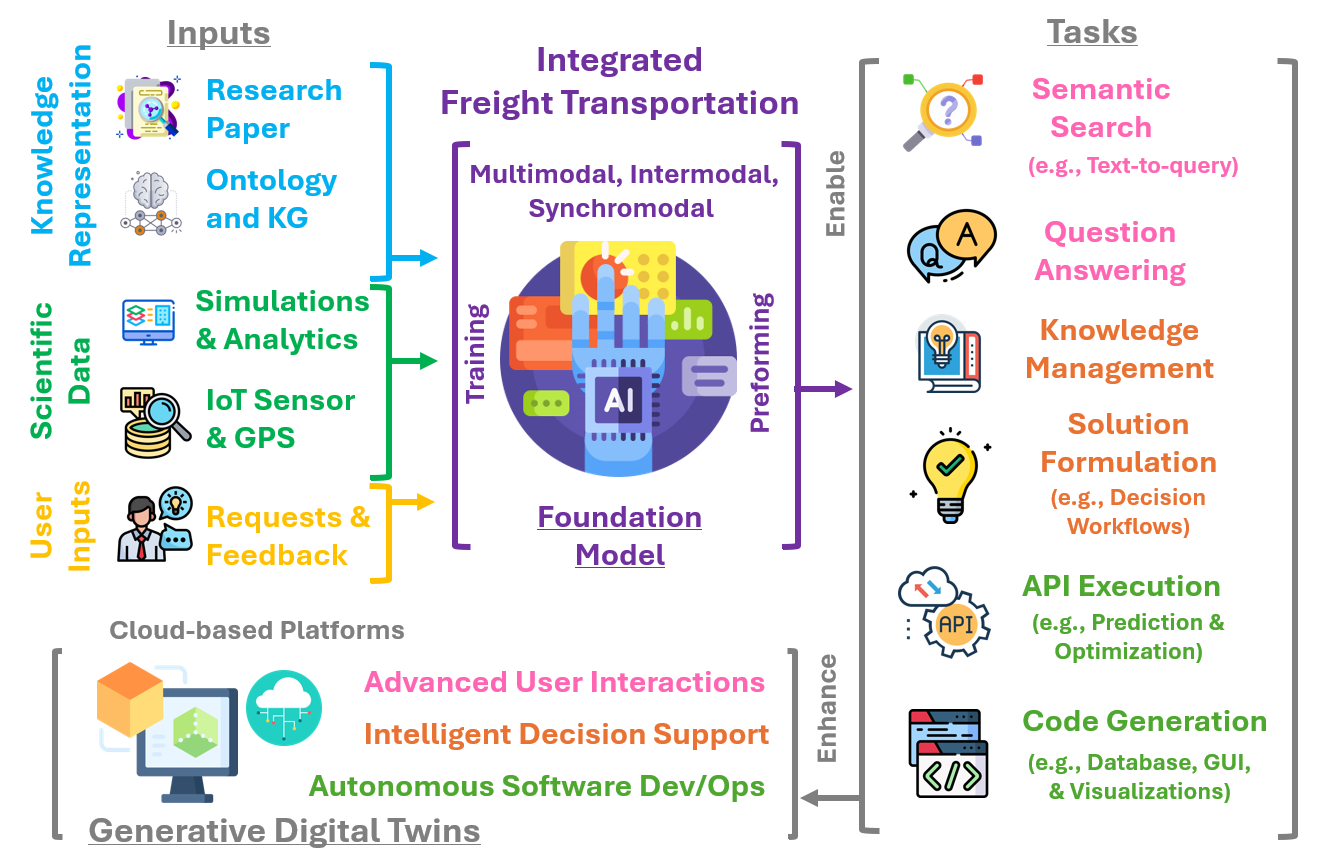}
    \caption{The concept of building a foundation model for optimizing integrated freight transpiration system.}
    \label{fig:concept}
\end{figure*}

The design and operation of these modern freight transportation paradigm, despite their efficiency, presents numerous methodological challenges that require comprehensive decision-making strategies to optimize transportation between specific Origin-Destination (OD) pairs. Effective approaches involve combining transportation modes to minimize delivery times, reduce costs, and lower GHG emissions while ensuring smooth transitions across diverse socio-economic environments \citep{aljadiri2023evaluating}. Achieving efficiency is hindered by the complexity and uncertainty of logistics, characterized by interconnected dynamics like movement, storage, deconsolidation, consolidation, and intermodal transitions \citep{lv2019operational, reis2019intermodal}. With the progression of the freight transport paradigm from intermodal to synchromodal, these complexities are further complicated by mode-specific constraints, geographic variations, and unpredictable factors such as traffic, extreme weather, and accidents\citep{demir2019green}. Developing a sustainable optimization strategy requires processing large volumes of data, conducting simulations, and employing sophisticated optimization algorithms to select the best combination of modes, routes, and practices from millions of alternatives \citep{archetti2022optimization}. This process is both complex and time-consuming, necessitating advanced solutions to address the dynamic nature of integrated freight systems.

In recent years, industry and academia have increasingly adopted digital twin approaches to enhance data- and simulation-driven processes for intermodal and synchromodal optimization. By leveraging advanced ICT and 5G networks, digital twins, virtual replicas of physical systems, allow stakeholders to monitor, simulate, and optimize freight transport operations in real time. Integrating traffic sensors, GPS data, freight simulations, and machine learning models, digital twins help predict disruptions, evaluate operational scenarios, and identify the most efficient strategies, thereby improving resilience, flexibility, and sustainability in supply chains \citep{busse2021towards, ambra2019digital}. Building on existing practices in urban digital twin development for situational awareness, predictive analytics, and decision support, this vision explores how emerging generative AI technologies can address the technical and methodological challenges in creating intelligent and efficient digital twins for intermodal freight systems. Inspired by many contemporary multimodal and multi-agent systems, this vision paper proposes a generative foundation model approach, powered by transformer-based language models, to automate data engineering and software development for prototyping digital twins. This work highlights an knowledge graph-driven approach for digital twin feature creation, analytics, and task automation towards a more intelligent, self-organizing, and general-purpose decision support tool (as depicted in Figure \ref{fig:concept}. We demonstrate initial results of this prototyping foundation model, using emerging embedding, vector databases, and retrieval-augmented generation (RAG) techniques. Additionally, we share insights on integrating these foundation models with conventional software engineering practices within a digital twin platform.

\section{DIGITAL TWINS FOR FREIGHT TRANSPORT}
This section begins by clarifying several similar concepts within the freight transportation system. Following this, it reviews existing digital twin applications in the freight transportation sector. Subsequently, recent advancements in generative AI technologies are examined, laying the groundwork for introducing the concept of foundation models and reviewing their applications in research and operational science. The aim is to define the relevant domain and technical concepts to establish a foundation for proposing the novel integration of foundation models into urban digital twin applications in later sections. 

\subsection{Concept Definition}
Before the review, we first define and clarify a few similar logistics paradigms in the freight transportation sector. Within the domain of logistics and supply chain management, multimodal, intermodal, and synchromodal freight transportation are distinct yet interconnected concepts. Multimodal freight transportation refers to the movement of goods using more than one mode of transportation (e.g., truck, rail, sea) under a single contract, but without the seamless integration between different modes, often involving manual intervention for transitions \citep{dua2019quality, rossolov2017research}. Intermodal freight transportation, on the other hand, is characterized by a more coordinated use of different transport modes, where goods are moved in standardized containers across multiple modes under a single journey, optimizing the efficiency of transitions between modes \citep{pencheva2022current}. Synchromodal freight transportation takes the concept further by integrating real-time data and dynamic decision-making, allowing the selection of the most suitable transport mode at each leg of the journey based on current conditions and customer requirements, offering greater flexibility and efficiency compared to the other methods \citep{ambra2019towards}. 

Other freight transportation paradigms include unimodal transportation (involving only one mode of transport), combined transport (specifically focusing on minimizing road usage by shifting to rail or inland waterways for long-distance segments), and co-modal transportation (promoting the efficient combination of transport modes without mode hierarchy) \citep{yang2024synchronizing, rossolov2017research}. Together, these paradigms represent the evolution of freight logistics toward increasingly integrated and adaptive systems to meet modern supply chain needs. As many of these paradigms are progression of the previous paradigm with extended capabilities, this paper proposes an integrated freight transportation paradigm, through the incorporation of the recent generative AI techniques and their exceptional knowledge extraction and reasoning capability, to enable an automated approach that can leverage the knowledge of previous paradigms to formulate adaptive solutions based on the nature and requirement of the real-world decision problems in large scale logistics system.  

\subsection{Freight Transportation Digital Twins}
This section reviews the development and application of digital twins across various freight transportation paradigms. We formulated the following Scopus query: TITLE-ABS-KEY((``digital twin" OR ``digital twins" OR ``digital twin technology") AND (``freight" OR ``cargo" OR ``logistics") AND (``intermodal" OR ``multimodal" OR ``synchromodal")), and supplemented it with additional searches using Google Scholar to identify studies that employ or develop digital twins for managing and optimizing freight transportation. These combined searches across academic databases yielded only 15 relevant articles based on topic relevance. Most digital twin research in the broader transportation sector focuses on developing intelligent transportation and smart mobility systems, often emphasizing human mobility and urban traffic dynamics \citep{xu2023smart, li2021emerging}. However, digital twin applications in the freight transportation sector remain scarce.

\subsubsection{Multimodel Freight Transportation}
\citet{busse2021towards} presents a Digital Supply Chain Twin (DSCT) as a comprehensive solution for optimizing complex multimodal supply chains. The DSCT integrates real-time data from logistics actors like freight forwarders, shipping companies, and terminal operators to provide visibility across the supply chain. It simulates scenarios for optimizing resource allocation, mitigating disruptions, and improving decision-making. Utilizing technologies such as IoT, 5G, cloud computing, and AI, the DSCT synchronizes different transport modes, improving efficiency, reducing lead times, and enhancing supply chain resilience and sustainability. \citet{issa2024railway} introduces a Digital Twin (DT) for railway systems to support multimodal freight transport. Developed at SNCF Réseau, the DT provides a real-time representation of railway infrastructure, facilitating synchronization with road, maritime, and inland navigation modes. Integrating a Service-Oriented Architecture (SOA) allows seamless data sharing among stakeholders such as infrastructure managers and logistics operators. This unified data source optimizes multimodal transport planning, enhances operational efficiency, promotes a modal shift to sustainable options like rail, and supports decarbonization efforts.
\citet{dorofeev2024improving} discusses the use of Digital Twins (DTs) in Transportation Management Systems (TMSs) for enhancing freight transport reliability. A Digital Twin of an Organization (DTO) is proposed, using ontological modeling and Process Mining to reconstruct business processes from event logs and identify deviations impacting reliability. Integrating data from GPS/GLONASS, IoT sensors, and TMS event logs, the DTO provides real-time insights into operations, facilitating proactive corrections, optimized route planning, and improved cargo safety, thereby enhancing overall transportation efficiency. 
\citet{sun2024enhancing} presents a Digital Twin (DT) to improve shipyard transportation efficiency via dynamic scheduling of transporters. The DT creates a real-time virtual model, integrating data from GIS, sensors, and physical transporters to optimize scheduling and address constraints like road restrictions and irregular block shapes. This continuous synchronization between physical and digital spaces enables proactive management, reduced idle times, and enhanced resource utilization, significantly improving shipbuilding operations and stakeholder coordination.

\subsubsection{Intermodal Transportation}
\citet{morra2019case} provides a comprehensive survey of DT technology applications in the railway sector, emphasizing the potential of DTs to revolutionize freight transportation systems. The Digital Twin for Railway (DTR) is highlighted as a virtual representation of physical railway assets, which can enhance infrastructure management, predictive maintenance, and operational efficiency. By integrating technologies like Internet of Things (IoT), Artificial Intelligence (AI), and Big Data, DTR allows for real-time monitoring and management of railway systems, enabling proactive interventions and reducing operational disruptions. The article points out that DTR is still in its early stages but holds promise for enhancing maintenance scheduling, structural health monitoring, and safety in railway logistics, ultimately supporting a more reliable and sustainable freight transportation system. This survey underscores the importance of DTs in making data-driven decisions, improving collaboration among stakeholders, and enabling the optimization of Intermodal freight networks.

Digital twin applications specifically designed for operating intermodal transportation systems are rare. Typically, the real-time data analytics and predictive capabilities of digital twins have been utilized to address synchromodal freight transportation, an intermodal challenge involving real-time decision support.

\subsubsection{Synchromodal Freight Transportation}
\citet{ambra2019digital} explores the use of the DT concept to enhance synchromodal freight transportation by reducing uncertainty and enabling more dynamic, flexible logistics operations. The DT acts as a virtual mirror of the physical transportation system, integrating data from various sources, including Geographic Information Systems (GIS), sensors, and agent-based models, to create a comprehensive and real-time representation of assets like trucks, trains, and barges. By continuously updating itself based on real-time data, the DT provides a dynamic platform to simulate and predict future system states, which can be used to assess the impact of disruptions and optimize transportation processes. The use of Monte Carlo simulations in the DT enables stakeholders to evaluate potential disruptions, plan re-routing, and assess the performance of different modal options under varying conditions. This integration of real and virtual systems allows the synchromodal freight network to adapt quickly to changes, improving its resilience, reducing lead times, and minimizing costs. The article emphasizes that the DT supports synchromodal decision-making by enhancing situational awareness, facilitating dynamic re-routing, and optimizing modal choices, ultimately contributing to a more efficient and sustainable freight transportation system. \citet{ambra2020agent} explores the use of Agent-Based Digital Twins (ABM-DT) to support synchromodal freight transportation by fusing virtual and physical environments for optimal decision-making. The Digital Twin (DT) serves as a virtual representation of the physical freight transport system, allowing for real-time integration of different transport modes such as road, rail, and inland waterways. The DT is built using agent-based modeling and Geographic Information Systems (GIS), enabling detailed and dynamic simulations of asset movements and transportation processes. The DT concept is applied to bridge the gap between real-time data from physical systems and their virtual counterparts. Real-time data feeds from assets, including vessels and trucks equipped with IoT and GPS, are fed into the digital environment, where they are represented as agents. This connection enables decision-makers to optimize route planning and adapt to disruptions dynamically, improving the overall flexibility and efficiency of the freight transportation system. The integration of these digital and physical elements aims to facilitate a modal shift from road to more sustainable modes like rail and inland waterways, contributing to environmental goals by reducing greenhouse gas emissions. The article demonstrates how the Digital Twin can enhance the decision-making process, provide better transparency, and make the synchromodal transport system more adaptive and resilient to changes.

\subsubsection{Knowledge Gap and Opportunities for Improvements}
Many existing digital twins adopt solid software engineering practices, such as Service-Oriented Architecture (SOA) and cloud computing, to host essential data and simulations (e.g., agent-based models), creating a virtual replica of real-world freight transportation systems. Despite their value and significance, these digital twins are developed using conventional digital twinning approaches that have several limitations. These limitations can be addressed through the recent advancements in generative AI technologies and the foundation model paradigm, paving the way for next-generation software development. 

\begin{description}
\item[Single-Purpose Applications:] Traditional digital twin initiatives are heavily rooted in conventional software engineering practices, employing dedicated data analytics, simulations, and optimization techniques within fixed computing environments. Consequently, these digital twins are typically developed as standalone products, tailored to perform specific, predefined functions (e.g., workflows and features) for targeted transportation paradigms, such as multimodal or intermodal freight systems. This approach introduces several limitations: the lack of interoperability with other systems and tools, limited adaptability to decision-making challenges across diverse freight paradigms, and restricted software extensibility. The integration of new methods, datasets, or research developments often demands significant time and effort. These constraints impede scalability and adaptability, highlighting the necessity for a more intelligent and automated approach—one that paves the way for general-purpose or multi-purpose digital twins capable of accommodating a range of freight transportation scenarios.
\item[Traditional User Interaction:] Existing digital twins for freight transportation predominantly utilize Graphical User Interfaces (GUIs) and predefined user workflows for interaction. These interfaces, while sufficient for executing specific tasks, offer limited flexibility and often require users to possess specialized domain knowledge. User interactions are restricted to a fixed set of buttons, menus, or dashboards, which inherently limits the breadth of queries and actions that can be performed. While effective for structured tasks, this rigid interface design falls short in supporting nuanced user needs or enabling complex, interactive decision-making. Recent advancements in conversational AI, such as chatbots, offer a more dynamic interaction model where users can communicate through natural language—be it text or voice—emulating a human conversation. This fosters open-ended inquiries, personalized responses, and deeper data exploration, empowering users to ask complex questions and receive contextually relevant answers. By bridging the gap between rigid system functionality and intuitive, human-like interactions, conversational AI enhances accessibility and usability, thereby offering a richer, more adaptive decision-making experience.
\item[Static Data Sources and Limited Dynamic Optimization:] Traditional digital twins, built upon conventional software engineering methodologies, generally depend on static, pre-defined datasets, which are inadequate for capturing the dynamic nature of transportation systems. This reliance on static data restricts the ability of these digital twins to support adaptive, real-time optimization and decision-making, as they cannot effectively respond to rapid changes or integrate new information beyond their initial dataset. Recent advancements in AI-driven software, particularly those employing Large Language Models (LLMs) and Retrieval-Augmented Generation (RAG) mechanisms, address this shortcoming by actively searching, retrieving, and utilizing constantly updating data and information available online. These AI-powered digital twins dynamically incorporate external knowledge—such as real-time traffic conditions, regulatory updates, or new research findings—into their existing models, resulting in more accurate analytics, simulations, and optimizations. This capacity for real-time adaptation significantly improves the flexibility and responsiveness of digital twins, transforming them into powerful tools for navigating the complexities of modern freight systems.
\end{description}

With the ongoing trends of developing AI agent and AI powered system to support scientific research, this paper propose a novel paradigm to revolutionize the process for creating digital twins for research by harassing the power of recent emerging generative AI models \citep{xu2024genai, xu2024leveraging}. 


\subsection{Raise of Generative AI and Foundation Models}
The recent evolution of generative AI models, such as GPT-3, GPT-4, DALL-E, and other transformer-based architectures, has paved the way for building foundation models that are transforming how research is conducted across different scientific disciplines \citep{yenduri2024gpt}. These generative AI models have demonstrated remarkable capabilities in understanding, generating, and synthesizing complex forms of data, whether textual, visual, or multimodal, thus giving rise to the concept of foundation models \citep{waqas2023revolutionizing}. These models are often large-scale pre-trained AI models that can serve as a versatile base across a wide array of specialized scientific tasks \citep{zhang2024urban}. The power of generative AI in natural language understanding, image synthesis, and data transformation showcases the potential of adapting these models to domain-specific applications. Foundation models harness this power by being trained on expansive, domain-specific datasets and knowledge, ranging from scientific literature, experimental observations, to multimodal research data \citep{chen2024evolution, bommasani2021opportunities}. This enables them to act as foundational frameworks that can be fine-tuned or utilized directly to perform critical tasks, including knowledge extraction, data analysis, automated hypothesis generation, simulation, and complex data integration.

In recent years, foundation models represent a shift from developing bespoke machine learning models for each scientific problem to leveraging a general-purpose model that can be adapted to a multitude of use cases within a given discipline \citep{bommasani2021opportunities}. They learn broad, transferable representations that can be specialized for niche applications, effectively functioning as a robust ``knowledge base" with an extensive understanding of the data landscape in fields such as medical, climate science, and urban science \citep{nguyen2023climax, chen2024evolution, moor2023foundation}. This has resulted in applications like BERT-based models for analyzing chemical compounds \citep{zhang2024scientific}, large-scale language models for synthesizing biomedical information \citep{tian2024opportunities}, and multimodal models that combine text, images, and structured data for complex interdisciplinary tasks in geoinformatics and remote sensing \citep{mai2024opportunities}. Foundation models in scientific research provide new levels of scalability, adaptability, and speed that were previously unattainable.

\subsection{Advantages of Foundation Models}
Compared to traditional simulation and data analysis methods, foundation models exhibit several notable advantages. Traditional methods, such as computational simulations or statistical modeling, often require substantial domain-specific expertise to define parameters, run models, and interpret results, which limits scalability. Furthermore, these models are typically task-specific and require structure data, making them challenging to adapt across different domains or datasets without significant manual reconfiguration and data engineering efforts. 

In contrast, foundation models are inherently generalizable, leveraging their extensive pre-training to seamlessly incorporate different types of data, including structured and unstructured, and respond to novel research questions with little adjustment \citep{myers2024foundation, zhang2024data}. These models can efficiently analyze vast amounts of complex data, generate predictions, conduct self-learning from the current research articles, interpreting charts and graphs, and even create new insights and hypotheses \citep{bommasani2021opportunities, huang2024pixels}. These hypothesis and insights often with a level of insight that is difficult for traditional approach to match due to their inflexibility and the intense computational resources they demand. By combining generative capabilities with cross-domain adaptability, foundation models open new avenues for scientific inquiry, accelerating discovery cycles and enabling the type of integrated, systems-level research required to address modern scientific challenges \citep{bommasani2021opportunities}. This evolution represents a fundamental change in scientific research, creating general-purpose and multi-purpose research applications that automate complex and time-consuming research tasks to accelerate the generation of knowledge and discoveries.

\section{METHODOLOGY}

\begin{figure*}[hbt!]    
\includegraphics[width=1\textwidth]{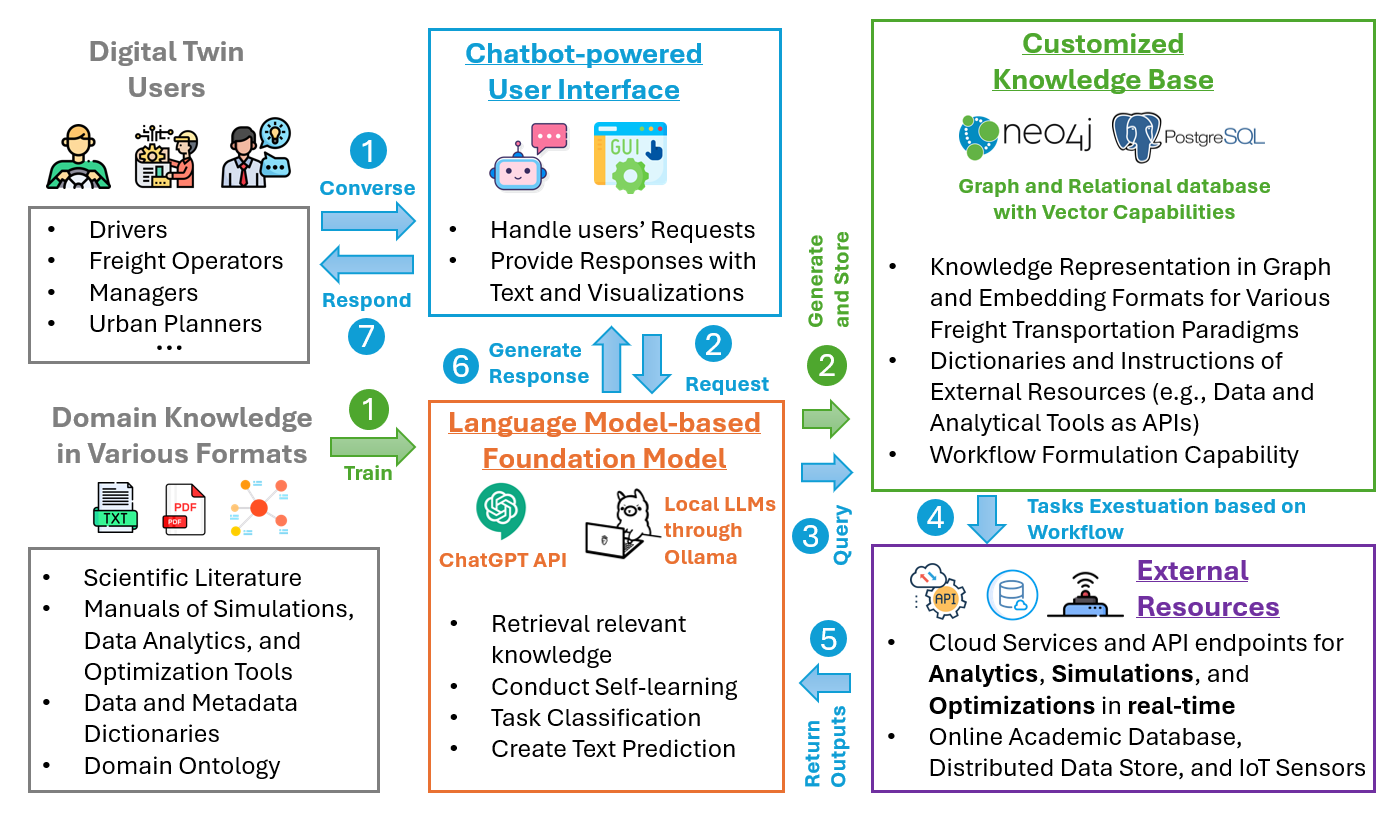}
    \caption{The technical framework for developing incorporating foundation models to enhance digital twins with two workflows (1) Knowledge Ingestion and Management (depicted with green arrows) and (2) General-purpose decision support for operations research (depicted with blue arrows).}
    \label{fig:workflow}
\end{figure*}

To develop a general-purpose foundation model that supports complex decision-making and optimization in the planning and operation of integrated freight transportation systems, we propose a technical framework that leverages cutting-edge transformer-based language models and vector databases through a Retrieval-Augmented Generation (RAG) paradigm, as depicted in Figure \ref{fig:workflow}). Our proposed framework consists of four major components that enable two primary workflows:
\begin{description}
    \item [Chatbot-powered user interface] is implemented using Generative Pre-trained Transformers, utilizing both the ChatGPT API and local instances of LLMs through the Python Ollama framework. The chatbot-powered interface enhances user interaction with the digital twin application, allowing users with varying technical backgrounds or no domain expertise to intuitively and effectively use the general-purpose digital twin for performing a wide range of specialized tasks.
    \item [Foundation Model] serves as the general-purpose problem-solving engine, driven by domain knowledge and data. It is developed using transformer-based language models implemented via local instances enabled by the Python Ollama framework. The model is designed with strong self-learning, reasoning, and self-organizing capabilities, allowing it to autonomously formulate solution workflows and execute specialized tasks using various research tools (e.g., statistical, mathematical, machine learning algorithms, domain simulations, and optimization tools). These research tools are available as cloud-based or traditional web services.     
    \item [Customized Knowledge Base] stores the most up-to-date representations of evolving research methodologies, tools, and datasets, which are published in various forms (e.g., journal articles, conference proceedings, and technical manuals). It is populated using the foundation model's natural language understanding capabilities to process large volumes of related text documents, leveraging language models to extract entities and relationships through text embeddings to generate knowledge graphs based on these contents. The knowledge graph is then stored in a Neo4j database with vector storage capabilities, containing fundamental information required to define complex decision support problems in logistics, as well as providing modular knowledge components for formulating science-based solutions. The rationale of the process have been proved by our previous studies \citep{xu2024automating, tupayachi2024towards}
    To extend the foundation model's basic question-answering abilities to autonomous task execution, which allows it to automatically conduct data queries, management, and analytics, as well as configure and execute domain-specific simulations (e.g., Freight Analysis Framework, spatial analytics) and optimization algorithms (e.g., AnyLogic, Gurobi), the knowledge base must also store data and metadata dictionaries, along with core instructions to guide the foundation model in performing these specialized tasks in a cyber-delivery manner. These dictionaries and instructions are stored in a PostgreSQL relational database with vector storage extensions.  
    \item [External Resources] consist of scientific data, simulation models, and research tools implemented using a cyber-delivery paradigm, which involves deploying scientific models as adaptive web services \citep{mutschke2011science, narayanan2003analysis}. Access to these services is documented in the knowledge base, allowing the foundation model to directly interface with cloud-based or traditional web-based API endpoints. In this way, the foundation model can autonomously execute specialized research tasks and utilize simulation and analytical outputs to generate quantitative responses that reflect the statistical and physical dynamics underlying complex urban logistics systems.
\end{description}

Building upon these components, the proposed framework enables two primary workflows. The first workflow involves training the foundation model using the most up-to-date knowledge on cutting-edge research and tools in the freight transportation domain, as highlighted by the green arrows in Figure \ref{fig:workflow}. The second workflow is responsible for responding to requests from various types of users (e.g., truck drivers, freight operators, managers, and urban planners) by leveraging the foundation model's reasoning and generative capabilities, as illustrated by the blue arrows in Figure \ref{fig:workflow}.

\section{PRELIMINARY RESULTS}
\begin{figure*}[hbt!]    
\includegraphics[width=1\textwidth]{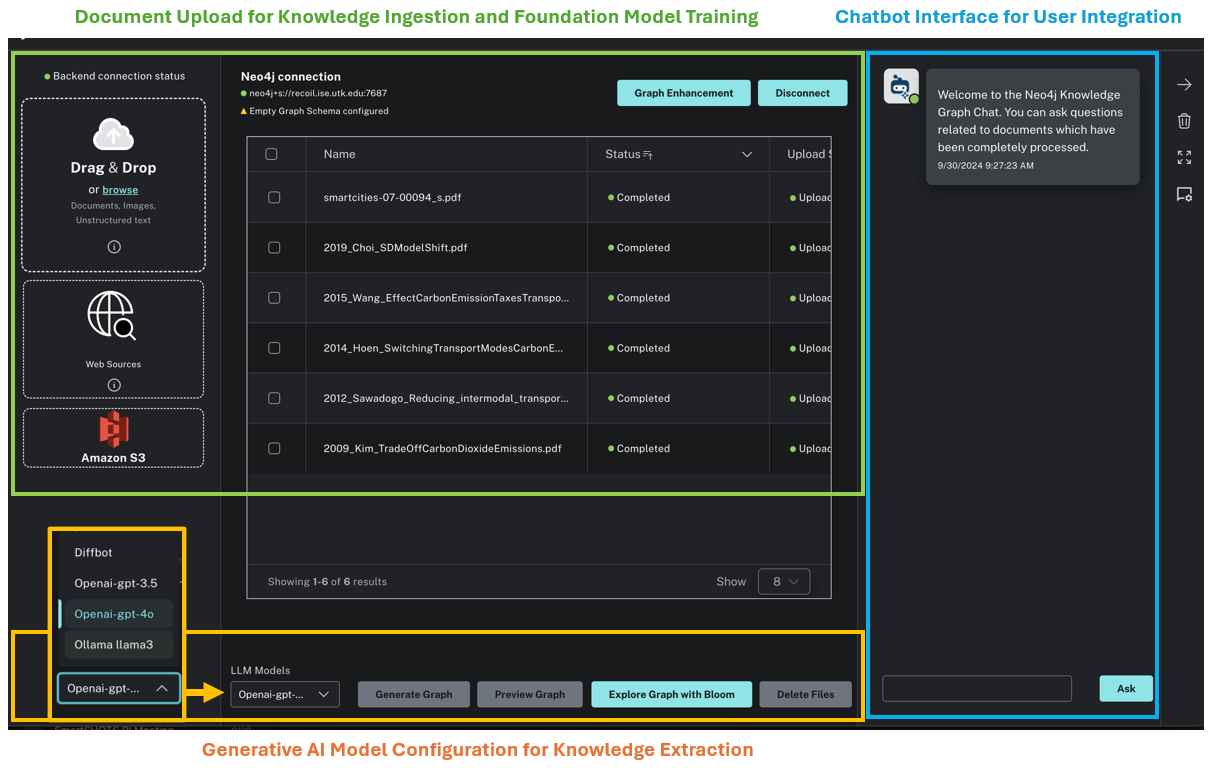}
    \caption{A Prototyping interface of the proposed foundation modal developed using open-source packages.}
    \label{fig:userintreface}
\end{figure*}

To evaluate the capability and performance of the proposed framework, we have implemented a prototype system that integrates all four core components using cutting-edge data science and software dismantlement technologies in a docker environment (as depicted in Figure \ref{fig:userintreface}). This prototype was designed to validate the feasibility of the technical framework in supporting complex decision-making and optimization tasks within the context of an integrated freight transportation system. The implementation allowed us to conduct a series of experiments to assess the system's ability to autonomously formulate solutions, interact with external data sources, and effectively respond to user requests, thereby demonstrating the practicality and potential of the framework in real-world applications. 

\subsection{Demonstration}
 
\begin{figure*}[hbt!]
    \includegraphics[width=1\textwidth]{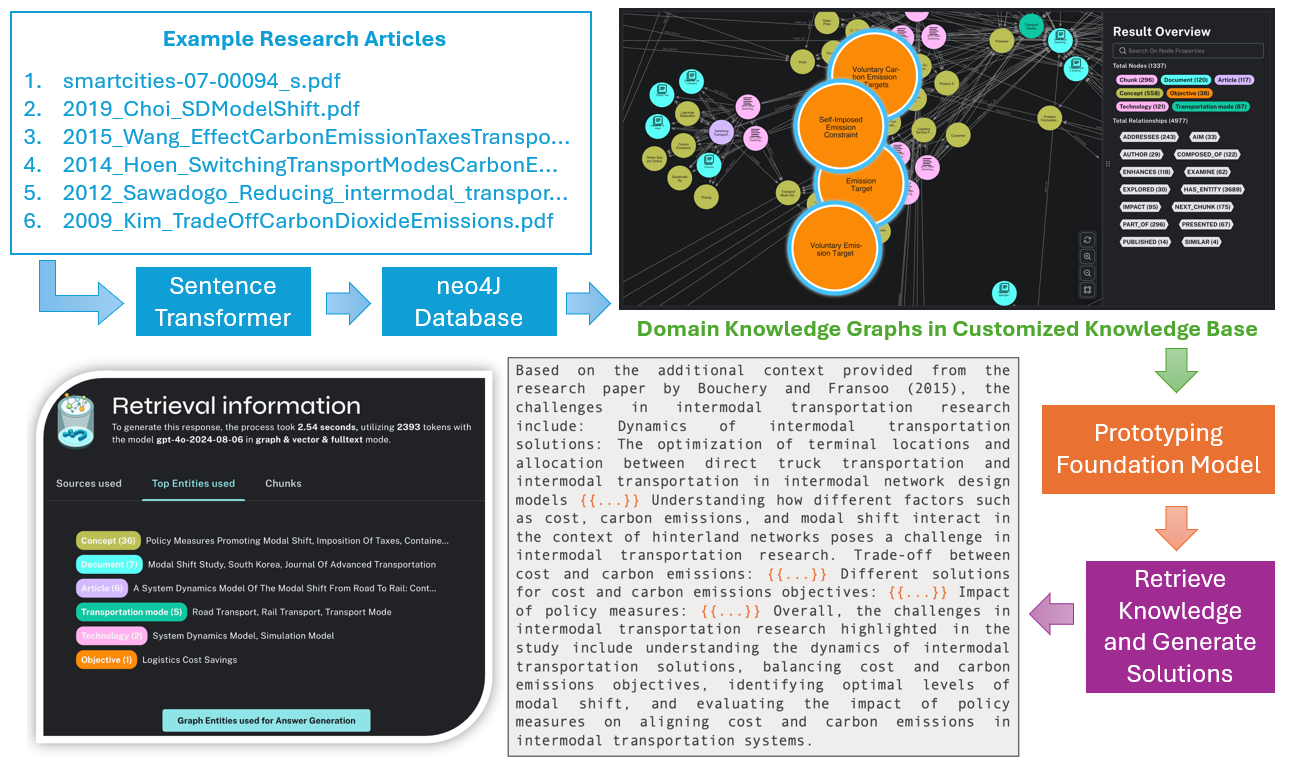}
    \caption{A demonstration of the prototype's learning and knowledge extraction capabilities, showcasing its ability to analyze contemporary intermodal research articles and apply the RAG paradigm. The generated knowledge graph is then utilized to provide solutions for user-defined requests.}
    \label{fig:demonstration}
\end{figure*}

Figure \ref{fig:demonstration} presents an showcase of the prototype for constructing a customized domain-specific knowledge base and utilizing it for solution generation in intermodal transportation research. The process begins with a set of research articles, which are analyzed using a Sentence Transformer to extract meaningful embeddings. These embeddings are stored in a Neo4j database to build a domain knowledge graph, visualized here with interconnected nodes representing various key concepts, objectives, technologies, and transportation modes. The knowledge graph serves as the foundation for retrieving information and identifying relationships between entities. The retrieval process provides information, as demonstrated in the retrieval dashboard, showing entities such as documents, concepts, and technologies. This context is used to develop a prototyping foundation model that leverages the domain knowledge graph to retrieve relevant knowledge and generate solutions for question asked related to the optimization of intermodal transportation system, such as balancing cost and carbon emissions, optimizing terminal allocation, and evaluating policy measures.

By expanding the scope of uploaded research articles to encompass other freight transportation paradigms, the prototype can generate customized, adaptive solutions tailored to address challenges in operating and optimizing multimodal and synchromodal freight systems. At the current stage of experimentation, we have implemented external resource components by deploying essential data analytics and simulation outputs as adaptive web services through API endpoints using the Python Django framework. These resources include data from the Freight Analysis Framework and the Freight and Fuel Transportation Optimization Tool.

The prototype has successfully demonstrated its ability to utilize the generated RAG-based solutions to access corresponding API endpoints and retrieve data and simulation outputs based on user-defined scenarios. Further experimentation is needed to assess the foundational model's capability to perform and coordinate a sequence of specialized tasks via API endpoints, as well as to integrate outputs from individual tasks into a cohesive workflow that provides insights and quantitative recommendations for optimizing complex freight transport systems. Additionally, the prototype's capability to handle, process, and analyze GIS data, such as routing through large road networks, requires further exploration and testing.

\section{VISION AND CONCLUSIONS}
The vision presented in this paper highlights the transformative potential of leveraging generative AI and foundation models to enhance digital twins for optimizing integrated freight transportation systems. Traditional digital twin applications, while effective in modeling and simulating complex systems, often face limitations due to their reliance on conventional software engineering practices. These practices are typically constrained by static data sources, fixed workflows, and limited adaptability, which restrict their ability to dynamically respond to evolving logistics challenges. By incorporating emerging generative AI technologies, particularly foundation models like transformer-based language models, we can overcome these limitations and create a new paradigm of intelligent, autonomous digital twins. These digital twins would not only offer real-time situational awareness but also provide adaptive optimization and decision-making capabilities, revolutionizing how urban logistics are managed.

Our proposed framework leverages the generative capabilities of foundation models to automate data engineering, knowledge discovery, and decision-making processes. This enables digital twins to evolve into versatile platforms capable of addressing diverse freight transportation scenarios, ranging from multimodal to synchromodal systems. By integrating RAG and knowledge graph-driven techniques, next-generation digital twins can autonomously enhance situational awareness, perform complex data analysis, and generate actionable insights. Our vision is to develop a general-purpose tool to support decision-making in integrated freight transportation systems, thereby significantly enhancing efficiency, flexibility, and sustainability. In future work, we plan to interface the foundation model with real-time traffic, weather, and freight transportation data, utilizing advanced research tools and software like AnyLogic and Gurobi for more sophisticated optimization tasks.

In conclusion, the integration of generative AI technologies and foundation models into digital twin platforms offers a promising path forward for advancing urban logistics and addressing the complexities inherent in modern freight transportation. The preliminary results of our prototype system demonstrate the feasibility of this approach in enhancing decision-making processes, promoting resilience, and supporting decarbonization efforts. Future work will focus on refining the framework, expanding its capabilities for handling diverse data sources, and optimizing its performance for large-scale deployment in real-world applications. By adopting these emerging technologies, we can create digital twins that are not only more intelligent but also capable of shaping a more sustainable future for freight transportation.

\section{ACKNOWLEDGEMENT}
This work was supported by the U.S. Department of Energy (DOE), Advanced Research Projects Agency–Energy (ARPA-E), under project DE-AR0001780. We extend our gratitude to our collaborators from the University of Tennessee, Knoxville. Additionally, several icons from www.flaticon.com were utilized in creating figures for this research.

\section{DISCLAIMER}
This manuscript has been authored by UT-Battelle, LLC, under contract DE-AC05-00OR22725 with the US Department of Energy (DOE). The US government retains and the publisher, by accepting the article for publication, acknowledges that the US government retains a nonexclusive, paid-up, irrevocable, worldwide license to publish or reproduce the published form of this manuscript, or allow others to do so, for US government purposes. DOE will provide public access to these results of federally sponsored research in accordance with the DOE Public Access Plan (http://energy.gov/downloads/doe-public-access-plan).

\small
\bibliographystyle{elsarticle-harv}
\bibliography{refs}

\end{document}